\documentclass[aps,preprint,amsmath,amssymb,superscriptaddress,nofootinbib]{revtex4}

\usepackage{graphicx}
\begin{document}

\title{Zero-mass limit of a Dirac spinor with general spin orientation}

\author{\.{I}. \c{S}ahin}
\email[]{inancsahin@ankara.edu.tr}
 \affiliation{Department of
Physics, Faculty of Sciences, Ankara University, 06100 Tandogan,
Ankara, Turkey}

\begin{abstract}
The helicity eigenstates which describe the fermions with a special
spin orientation (parallel or antiparallel to the direction of
momentum) provide considerable simplification in calculations.
Hence, it is generally preferred to use helicity basis during the
calculations in Relativistic Quantum Mechanics or Quantum Field
Theory. Possibly because of the above reason, the Dirac spinors
describing a general spin orientation have been ignored in many
textbooks. Although the helicity eigenstates give almost complete
understanding of the behavior of the free Dirac solutions, the
zero-mass limit is one of its exception. The zero-mass behavior of
the free spinor with general spin orientation and its relation with
chirality eigenstates has been skipped in textbooks and hence it
deserves a clear, detailed investigation. In this paper we obtain
the free Dirac spinors describing a general spin orientation and
examine their zero-mass limit. We also briefly discuss some of the
implications of this small-mass behavior of the spinors on particle
physics.

\end{abstract}


\maketitle

\section{Introduction}

Chirality is defined by the operator Dirac matrix $\gamma^5$. Its
eigenstates with eigenvalues +1 and -1 are sometimes called right
and left-handed or positive and negative chirality eigenstates.
Chirality has a fundamental importance in the Standard Model of
particle physics. Weak interactions behave differently for different
chirality fermion fields \cite{Quang}. Although chirality is a
Lorentz invariant quantity it is not conserved for massive fermions.
This is obvious from the fact that the chirality operator does not
commute with the mass term in the Dirac Hamiltonian. On the other
hand, helicity which is simply defined as the projection of the spin
onto the direction of momentum \footnote{If a fermion is polarized
in the direction parallel (antiparallel) to the momentum then it is
described by a positive or right-handed helicity (negative or
left-handed helicity) state.}, is a conserved quantity but not
Lorentz invariant. Therefore,  in this perspective chirality and
helicity have somewhat opposite characteristics for massive fermions
\cite{Pal:2010ih}. However, for massless fermions they become both
conserved and Lorentz invariant.

It is well known that the helicity and the chirality eigenstates
coincide in the zero-mass limit \cite{Kim,Ryder,GreinerRQM}. On the
other hand, a general spin state does not exhibit such a behavior.
For instance, the spinor for a free fermion which is transversely
polarized relative to the direction of its momentum is always given
by the same superposition $u^{(T)}=\frac{1}{\sqrt
2}(u^{(+)}+u^{(-)})$ of positive and negative helicity spinors no
matter how small the mass parameter is. Since the positive and
negative helicity spinors converge to positive and negative
chirality eigenstates for $m\to0$, the transversely polarized spinor
$u^{(T)}$ is always given by a mixed chirality eigenstate and hence
does not converge to one of the chirality eigenstates left-handed or
right-handed even for infinitesimal values of the mass. Thus we
conclude that the zero-mass limit of the free Dirac spinor with
arbitrary spin orientation does not necessarily result in a
chirality eigenstate. Although the solutions of the free Dirac
equation which describe noninteracting spin-1/2 particles such as
electrons have been widely studied in many textbooks, the free
solutions describing a general spin orientation have been commonly
omitted. Many of the books on this subject, consider only a special
type of Dirac solutions: the so-called helicity states. Only in few
of the books that study the Dirac equation, the authors implicitly
demonstrate the methodology which can be used to obtain the free
solutions for general spin states but the explicit expressions are
not given. We can give the references \cite{Kim,Ryder,GreinerRQM} as
an example in this respect. On the other hand, the spinors
describing general spin orientations are very essential to
understand the ultrarelativistic/zero-mass behavior of the free
Dirac solutions. As far as we know this fact has been ignored in all
textbooks, and ultrarelativistic/zero-mass limit has been analyzed
using only the special solutions namely helicity states. The main
purpose of this paper is to fill this gap. Our another purpose is to
deliver a better understanding of the spin phenomena in Relativistic
Quantum Mechanics. The comparison of two different methods given in
sections II and III makes the concepts clearer. The second method
demonstrated in section III provides a new simple way to obtain free
spinors with general spin. The paper is essentially prepared for
educational purposes and addresses graduate level students but it is
also useful for undergraduates who have attended courses on
Relativity and Quantum Mechanics.

In the following sections, we are going to obtain Dirac spinors
describing a general spin orientation using two different methods
and then analyze their zero-mass behavior. In section II the spinors
describing a general spin orientation are going to be obtained by
applying Lorentz transformations to a spinor at rest. In section III
the spinors for a general spin are going to be obtained with the
help of the covariant spin projection operator. In the discussion
section (section IV) we are going to examine the zero-mass behavior
of a spinor with general spin orientation and discuss briefly some
of its implications.

\section{Derivation of the spinor for general spin via Lorentz boost}

The orientation of spin relative to the direction of momentum is a
reference frame-dependent quantity for massive fermions. For
instance, we can perform a Lorentz transformation into a reference
frame in which the momentum of the particle is reversed. This
transformation obviously flips the helicity. One way to derive a
Dirac spinor describing a general spin orientation is to use Lorentz
transformations. In this method, the spinor for a moving fermion is
obtained by applying a Lorentz boost to the spinor in the rest frame
of the particle \cite{Kim,Ryder,GreinerRQM,Lerner}. In the rest
frame we can use the prescription of non-relativistic quantum
mechanics to define a spinor with general spin. By means of Lorentz
transformations we can easily get the relativistic description of a
general spinor, i.e., free solutions of the Dirac equation
describing a general spin orientation.

The Lorentz transformations of the Dirac spinors can be studied by
the Lie group $SL(2,C)$. The generators of $SL(2,C)$ can be written
in the following $4\times4$ matrix form\footnote{In this paper we
work in the Weyl representation of the Dirac matrices and use
natural $\hbar=c=1$ unit system.} \cite{Kim,Ryder}:
\begin{eqnarray}
J_m=\left(
      \begin{array}{cc}
        \frac {\sigma_m}{2} & 0 \\
        0 & \frac {\sigma_m}{2} \\
      \end{array}
    \right),\;\;\;\;
    K_m=\left(
    \begin{array}{cc}
        \frac {i\sigma_m}{2} & 0 \\
        0 & \frac {-i\sigma_m}{2} \\
      \end{array}
    \right)
\end{eqnarray}
where $\sigma_m$, $m=1,2,3$ are Pauli spin matrices. The generators
$J_m$ and $K_m$ are responsible for spatial rotations and Lorentz
boosts respectively. Without loss of generality, let us consider a
pure Lorentz boost along $z$-axis. The group element corresponding
to this transformation is given by \cite{Kim}
\begin{eqnarray}
\exp{(-i\xi K_3)}=\left(
                    \begin{array}{cc}
                      \exp{(\frac {\xi \sigma_3} {2})} & 0 \\
                      0 & \exp{(-\frac {\xi \sigma_3} {2})} \\
                    \end{array}
                  \right).
\end{eqnarray}
Here, $\xi$ is the rapidity parameter which is related to the
$\beta$ parameter of the Lorentz transformation as $\tanh
{\xi}=\beta$. It is easy to deduce the following identities from the
expansion of exponential functions and using the identities
$\left(\sigma_3\right)^{2k}=\mathbb{I}$,
$\left(\sigma_3\right)^{2k+1}=\sigma_3$.
\begin{eqnarray}
\exp{(\frac {\xi \sigma_3}
{2})}=\cosh{(\frac{\xi}{2})}\mathbb{I}+\sinh{(\frac{\xi}{2})}
\sigma_3 \\
\exp{(-\frac {\xi \sigma_3}
{2})}=\cosh{(\frac{\xi}{2})}\mathbb{I}-\sinh{(\frac{\xi}{2})}
\sigma_3
\end{eqnarray}
Hence the Lorentz boost along the $z$-axis can be written as
\begin{eqnarray}
\label{zboost}
 \exp{(-i\xi K_3)}=\left(
                    \begin{array}{cc}
                      \cosh{(\frac{\xi}{2})}\mathbb{I}+\sinh{(\frac{\xi}{2})}
\sigma_3 & 0 \\
                      0 & \cosh{(\frac{\xi}{2})}\mathbb{I}-\sinh{(\frac{\xi}{2})}
\sigma_3 \\
                    \end{array}
                  \right).
\end{eqnarray}
Assume that $S$ is the rest frame of the spin-1/2 fermion defined by
the coordinates $x^\mu=(x,y,z,t)$. In the rest frame we can use
Pauli spinors to define the spin of the fermion. Let $\vec{n}$ be
the unit vector in direction of the spin quantization axis in the
rest frame $S$. Then the non-relativistic $2\times2$ spin matrix is
$S_{\vec n}=\frac{1}{2}(\vec n\cdot \vec \sigma)$. Without loss of
generality, choose $\vec{n}=\sin\theta\; \hat x+ \cos\theta\; \hat
z$, i.e., $\vec{n}$ is in the $z$-$x$ plane which makes an angle
$\theta$ (polar angle) with respect to the $z$-axis. We do not loose
generality if we choose the azimuthal angle zero since we are going
to boost along the $z$-axis and the angle between the direction of
the boost and the unit vector $\vec{n}$ is the polar angle $\theta$.
Therefore, this particular choice does not change the general
results. The eigenvectors of $S_{\vec n}=\frac{1}{2}(\sin\theta\;
\sigma_x+\cos\theta\; \sigma_z)$ are
\begin{eqnarray}
\label{chi} \chi_+=\left(
           \begin{array}{c}
              \cos\theta/2\\
              \sin\theta/2\\
           \end{array}
         \right), \;\;\;\;\;\;\;\;
         \chi_-=\left(
           \begin{array}{c}
              -\sin\theta/2\\
              \cos\theta/2\\
           \end{array}
         \right).
\end{eqnarray}
Here $\chi_+$ and $\chi_-$ are the eigenvectors which correspond to
eigenvalues $\lambda=+1$ and $\lambda=-1$ respectively, i.e.,
$S_{\vec n}\;\chi_+=\chi_+$ and $S_{\vec n}\;\chi_-=-\chi_-$.
Sometimes the eigenvectors $\chi_+$ and $\chi_-$ are called
"spin-up" and "spin-down" states. We will use the notation
$\uparrow$ and $\downarrow$ for spin up and spin down. Up to now, we
have considered only 2-component spinors. In the Weyl
representation, 4-component spin-up and spin-down Dirac spinors can
be written as
\begin{eqnarray}
\label{restspinors} u^{(\uparrow)}_{RF} =\frac{1}{\sqrt 2}\left(
\begin{array}{c}
\chi_+ \\
\chi_+ \\
\end{array}
\right), \;\;\;\;\;\;\;\; u^{(\downarrow)}_{RF} =\frac{1}{\sqrt
2}\left(
\begin{array}{c}
\chi_- \\
\chi_- \\
\end{array}
\right)
\end{eqnarray}
where subscripts "RF" are used to stress that the spinors are
defined in the rest frame. Here, the factor $1/\sqrt 2$ comes from
the normalization $\bar u^{(r)} u^{(s)}=\delta^{rs}$. The spinors
for a moving fermion (in the $\pm z$ direction) can be obtained by
applying the Lorentz boost (\ref{zboost}) to the rest spinors:
\begin{eqnarray}
\label{movingspinors} u^{(\uparrow,\downarrow)}(p)=\exp{(-i\xi
K_3)}\;u^{(\uparrow,\downarrow)}_{RF}=\frac{1}{\sqrt
2}\left(\begin{array}{c}
\left[\cosh{(\frac{\xi}{2})}\mathbb{I}+\sinh{(\frac{\xi}{2})}
\sigma_3\right]\chi_{+,-} \\
\left[\cosh{(\frac{\xi}{2})}\mathbb{I}-\sinh{(\frac{\xi}{2})}
\sigma_3\right]\chi_{+,-} \\
\end{array}
\right)
\end{eqnarray}
After some algebra it is easy to show that
\begin{eqnarray}
\label{generalspinors1} u^{(\uparrow)}(p)=\frac{1}{\sqrt 2}\left(
                                     \begin{array}{c}
                                       \cos{\frac{\theta}{2}}\; e^{\xi/2} \\
                                       \sin{\frac{\theta}{2}}\; e^{-\xi/2} \\
                                       \cos{\frac{\theta}{2}}\; e^{-\xi/2} \\
                                       \sin{\frac{\theta}{2}}\; e^{\xi/2} \\
                                     \end{array}
                                   \right), \;\;\;\;\;\;\;\;
u^{(\downarrow)}(p)=\frac{1}{\sqrt 2}\left(
                                     \begin{array}{c}
                                       -\sin{\frac{\theta}{2}}\; e^{\xi/2} \\
                                       \cos{\frac{\theta}{2}}\; e^{-\xi/2} \\
                                       -\sin{\frac{\theta}{2}}\; e^{-\xi/2} \\
                                       \cos{\frac{\theta}{2}}\; e^{\xi/2} \\
                                     \end{array}
                                   \right).
\end{eqnarray}
The above spinors are given as a function of the rapidity $\xi$. One
can obtain the spinors in (\ref{generalspinors1}) as a function of
the energy and the momentum using the equality
$e^{\xi/2}=\left(\frac{E+|\vec p|}{E-|\vec p|}\right)^{1/4}$. This
equality is evident from the following definitions of the rapidity
parameter: $\sinh\xi=\gamma\beta$ and $\cosh\xi=\gamma$. In the
$c=1$ unit system the identities $\gamma=E/m$ and $\gamma\beta=|\vec
p|/m$ hold.

Let us discuss some special spin orientations. One of the important
special type of Dirac solutions are the helicity eigenstates. The
helicity operator ($\hat \Lambda$) is defined by the following
formula \cite{GreinerRQM}
\begin{eqnarray}
\label{helicityoperator} \hat \Lambda=\vec S\cdot\frac{\vec p}{|\vec
p|}
\end{eqnarray}
where $\vec S$ is the $4\times4$ spin matrix defined by
\begin{eqnarray}
\label{spinoperator} \vec S=\frac{1}{2}\left(
                              \begin{array}{cc}
                                \vec\sigma & 0 \\
                                0 & \vec\sigma \\
                              \end{array}
                            \right).
\end{eqnarray}
It is obvious from (\ref{helicityoperator}) that the helicity is
simply the projection of the spin onto the direction of momentum.
The solutions of the free Dirac equation which satisfy the
eigenvalue equations $\hat \Lambda u^{(+)}(p)=u^{(+)}(p)$ and $\hat
\Lambda u^{(-)}(p)=-u^{(-)}(p)$ are called the helicity eigenstates.
We observe from these equations that $u^{(+)}(p)$ and $u^{(-)}(p)$
are the eigenvectors of the helicity operator with eigenvalues $+1$
and $-1$ respectively. The helicity operator has no other
eigenvalues than $\pm1$. The helicity eigenstates can be obtained by
the choices $\theta=0$ and $\theta=\pi$ in (\ref{generalspinors1}).
According to the notation used in this paper, spin-up ($\uparrow$)
and spin-down ($\downarrow$) spinors given in
(\ref{generalspinors1}) correspond to positive ($+$) and negative
($-$) helicity spinors respectively if we choose $\theta=0$. We
observe from (\ref{generalspinors1}) that the spinors describing a
general spin can be written by the following superposition of the
helicity eigenstates \cite{Sahin:2015ofl}:
\begin{eqnarray}
\label{spinup}
u^{(\uparrow)}(p)=\cos{\left(\frac{\theta}{2}\right)}\;u^{(+)}(p)+\sin{\left(\frac{\theta}{2}\right)}\;u^{(-)}(p)\\
\label{spindown}
u^{(\downarrow)}(p)=\cos{\left(\frac{\theta}{2}\right)}\;u^{(-)}(p)-\sin{\left(\frac{\theta}{2}\right)}\;u^{(+)}(p)
\end{eqnarray}
Another interesting special orientation is the transverse
polarization which corresponds to the choice $\theta=\pi/2$. We see
from Eqs. (\ref{spinup}) and (\ref{spindown}) that the spinors for
transverse polarization are mixed states composed of helicity
eigenstates where each helicity eigenstate has equal probability.

The general spinors for antifermions can also be obtained in a
similar way. Here one should be careful about the fact that opposite
to the fermion case spin-up and spin-down states for antifermions
correspond to eigenvalues $\lambda=-1$ and $\lambda=+1$ of the
non-relativistic spin matrix $S_{\vec n}$ respectively (see the
paragraph below Eq.(\ref{chi})). Moreover there is a relative minus
sign between upper and lower 2-component subspinors of the
4-component spinor.

\section{Derivation of the spinor for general spin via covariant spin projection operator}

Dirac spinors for general spin can also be obtained by means of the
covariant spin projection operator. The covariant spin projection
operator is the covariant generalization of the non-relativistic
spin projection operator. Hence it is a reference frame independent
quantity and valid in any reference frame. Its explicit form is
defined as \cite{GreinerRQM,GreinerQED}
\begin{eqnarray}
\hat{\Sigma}(s)\equiv\frac{1}{2}(1+\gamma_5 \gamma_\mu s^\mu)
\end{eqnarray}
where $s^\mu$ is the spin-four vector which is the Lorentz
transformation of the rest vector $\left(s^\mu\right)_{RF}=(0,\vec
n)$ into a reference frame where the particle has a momentum
$p^\mu=(E,\vec p)$. Therefore, if the fermion is polarized in the
direction $\vec n$ in its rest frame then its spin-four vector can
be obtained through the Lorentz transformation $s^\mu=L^\mu_\nu
\left(s^\nu\right)_{RF}$. It is easy to show that the spin-four
vector can be written in the following form
\cite{GreinerRQM,GreinerQED}:
\begin{eqnarray}
\label{fourspin}
s^\mu=\left(\frac{\vec{p}\cdot\vec{n}}{m},\vec{n}+\frac{\vec{p}\cdot\vec{n}}{m(E+m)}\vec{p}\right).
\end{eqnarray}
The three-vector $\vec n$ is sometimes called the spin three-vector
and denoted by $\vec{s'}$. Here the prime symbol is used to indicate
that $\vec{s'}$ is not the spatial component of $s^\mu$, but they
are defined in different reference frames. Let $u^{(s)}(p)$ be the
spinor which describes a fermion with spin-four vector $s^\mu$. Then
in the rest frame of the particle its spin is quantized in the
direction $\vec n$. According to our notation, the sign of the
superscript $s$ represents the sign of the spin three-vector $\vec
n$. Therefore $u^{(-s)}(p)$ describes a fermion which is quantized
in the direction $-\vec n$ in the rest frame of the particle.
Obviously $+s$ ($-s$) corresponds to a spin-up (spin-down) state.
The orthogonality of spin-up and spin-down states can be written in
this notation as: $\bar u^{(s)}(p)u^{(-s)}(p)=0$. The covariant spin
projection operator $\hat{\Sigma}(s)$ projects to a polarized state
which is described by the spin-four vector $s^\mu$. Hence it has the
following properties:
\begin{eqnarray}
\hat{\Sigma}(s)u^{(s)}(p)=u^{(s)}(p),\;\;\;\;\;\;\;\;\hat{\Sigma}(s)u^{(-s)}(p)=0
\end{eqnarray}
Here we should note that the covariant spin projection operator not
only performs a projection into a polarized particle state but also
it performs a projection into a polarized antiparticle state.
Therefore similar equations can also be written for antispinors
$v^{(\pm s)}(p)$. Let us choose $\vec{n}=\sin\theta\; \hat x+
\cos\theta\; \hat z$. With this choice $\hat{\Sigma}(s)$ performs a
projection into a state with general spin orientation (Again without
loss of generality we assume that the azimuthal angle is zero and
$\vec n$ lies in the $z-x$ plane.). The desired spinor $u^{(s)}(p)$
can be obtained by applying $\hat{\Sigma}(s)$ to a subspace spanned
by the set of particle states
$\left\{|u^{(+s)}(p)>,|u^{(-s)}(p)>\right\}$. The projection
operator onto this subspace is given by $\sum_{s'} u^{(s')}(p) \bar
u^{(s')}(p)=u^{(+s)}(p) \bar u^{(+s)}(p)+u^{(-s)}(p) \bar
u^{(-s)}(p)$. This projection operator is independent from the
choice of $\vec n$ since for every fixed spin axis the spin-up and
the spin-down spinors span the same subspace and constitute its
basis. Its explicit form can be found in standard textbooks
\cite{Ryder,GreinerRQM} and given by
\begin{eqnarray}
\label{completenessrelation} \sum_{s'} u^{(s')}(p) \bar
u^{(s')}(p)=\frac{p^\mu\gamma_\mu+m}{2m}.
\end{eqnarray}
Eq.(\ref{completenessrelation}) is sometimes called the completeness
relation. The completeness relation implicitly contains the set of
states that we are looking for. If we apply $\hat{\Sigma}(s)$ from
the left and $\hat{\bar \Sigma}(s)=\hat{\Sigma}(s)$ \footnote{It can
be easily shown that $\bar u^{(s)}(p)=\bar
u^{(s)}(p)\frac{1}{2}(1+\gamma_5 \gamma_\mu s^\mu)$ where $\bar
u=u^{\dagger} \gamma_0$. Thus the bar "$-$" operation does not
change the form of the spin projection operator.} from the right to
the completeness relation (\ref{completenessrelation}) we obtain
\begin{eqnarray}
\label{completenessprojection} u^{(s)}(p) \bar
u^{(s)}(p)=\hat{\Sigma}(s)\left(\frac{p^\mu\gamma_\mu+m}{2m}\right)\hat{\Sigma}(s).
\end{eqnarray}
Here the projection ensures that the sum over $s'$ yields just one
term with $s'=s$. Equation (\ref{completenessprojection}) gives a
set of algebraic equations for the components of $u^{(s)}(p)$. The
right-hand side of Eq.(\ref{completenessprojection}) can be
calculated using the explicit expressions for Dirac matrices. For
simplicity we assume that the fermion is moving along the $z$-axis.
Then the right-hand side of Eq.(\ref{completenessprojection}) gives
a real $4\times4$ matrix. Hence we can choose the components of the
$u^{(s)}(p)$ spinor all real, $(u^{(s)}(p))_i=\alpha_i,\;\alpha_i
\in\Re,\;\; i=1,2,3,4$. After some algebra we obtain the following
matrix equation:
\begin{eqnarray}
\label{equation}
 \left(
  \begin{array}{c}
    \alpha_1 \\
    \alpha_2 \\
    \alpha_3 \\
    \alpha_4 \\
  \end{array}
\right)_{4\times1} \left(
    \begin{array}{cccc}
      \alpha_3 & \alpha_4 & \alpha_1 & \alpha_2 \\
    \end{array}
  \right)_{1\times4}=
\left(
  \begin{array}{cccc}
    \frac{1}{2}\cos^2(\frac{\theta}{2}) &\frac{(E+|\vec p|)\sin\theta}{4m}  &\frac{(E+|\vec p|)\cos^2(\frac{\theta}{2})}{2m}  &\frac{1}{4}\sin\theta  \\
    \frac{(E-|\vec p|)\sin\theta}{4m} &\frac{1}{2}\sin^2(\frac{\theta}{2})  &\frac{1}{4}\sin\theta  &\frac{(E-|\vec p|)\sin^2(\frac{\theta}{2})}{2m}  \\
    \frac{(E-|\vec p|)\cos^2(\frac{\theta}{2})}{2m} &\frac{1}{4}\sin\theta  &\frac{1}{2}\cos^2(\frac{\theta}{2})  &\frac{(E-|\vec p|)\sin\theta}{4m}  \\
    \frac{1}{4}\sin\theta &\frac{(E+|\vec p|)\sin^2(\frac{\theta}{2})}{2m}  &\frac{(E+|\vec p|)\sin\theta}{4m}  &\frac{1}{2}\sin^2(\frac{\theta}{2})  \\
  \end{array}
\right)_{4\times4}
\end{eqnarray}
The solution of the matrix equation (\ref{equation}) gives the
components of the spin-up spinor $u^{(\uparrow)}(p)$ for general
spin orientation. Similarly the components of spin-down spinor
$u^{(\downarrow)}(p)$ can be solved from (\ref{equation}) with the
replacement $\theta\to\theta+\pi$. This is obvious because spin-down
state corresponds to the orientation $-\vec{n}=-\sin\theta\; \hat x-
\cos\theta\; \hat z$ in the rest frame of the particle. We obtain
the following spinors from the solution of Eq.(\ref{equation}):
\begin{eqnarray}
\label{generalspinors2} u^{(\uparrow)}(p)=\frac{1}{\sqrt 2}\left(
                                     \begin{array}{c}
                                       \cos{\frac{\theta}{2}}\; \left(\frac{E+|\vec p|}{E-|\vec p|}\right)^{1/4} \\
                                       \sin{\frac{\theta}{2}}\; \left(\frac{E-|\vec p|}{E+|\vec p|}\right)^{1/4} \\
                                       \cos{\frac{\theta}{2}}\; \left(\frac{E-|\vec p|}{E+|\vec p|}\right)^{1/4} \\
                                       \sin{\frac{\theta}{2}}\; \left(\frac{E+|\vec p|}{E-|\vec p|}\right)^{1/4} \\
                                     \end{array}
                                   \right), \;\;\;\;\;\;\;\;
u^{(\downarrow)}(p)=\frac{1}{\sqrt 2}\left(
                                     \begin{array}{c}
                                       -\sin{\frac{\theta}{2}}\; \left(\frac{E+|\vec p|}{E-|\vec p|}\right)^{1/4} \\
                                       \cos{\frac{\theta}{2}}\; \left(\frac{E-|\vec p|}{E+|\vec p|}\right)^{1/4} \\
                                       -\sin{\frac{\theta}{2}}\; \left(\frac{E-|\vec p|}{E+|\vec p|}\right)^{1/4} \\
                                       \cos{\frac{\theta}{2}}\; \left(\frac{E+|\vec p|}{E-|\vec p|}\right)^{1/4} \\
                                     \end{array}
                                   \right).
\end{eqnarray}
The above spinors exactly coincide with the spinors
(\ref{generalspinors1}) obtained from Lorentz transformation.

The same method can be applied in the case of antifermions. However,
one should use the completeness relation $\sum_{s'} v^{(s')}(p) \bar
v^{(s')}(p)=\frac{p^\mu\gamma_\mu-m}{2m}$  for antifermions instead
of (\ref{completenessrelation}). Furthermore the definition of
spin-up and spin-down orientations should be reversed compare to the
fermion case.

\section{Discussion}

Let us examine the zero-mass limit of the spinors given in
(\ref{generalspinors1}) or (\ref{generalspinors2}). The zero-mass
limit corresponds to ultrarelativistic limit $\xi\to \infty$ or
equivalently $|\vec p|\to E$. We observe from
Eq.(\ref{generalspinors1}) that the terms proportional to
$e^{\xi/2}$ diverge in the zero-mass limit. Hence the spinors given
in (\ref{generalspinors1}) or (\ref{generalspinors2}) become
undefined in this limit. Therefore we should renormalize these
spinors to a finite value. If we take the zero-mass limit of the
spinors describing a general spin orientation (spinors in
(\ref{generalspinors1}) or (\ref{generalspinors2})) and normalize
them according to the formula $u^{\dag(r)} u^{(s)}=\delta^{rs}$, we
get the following spinors:
\begin{eqnarray}
\label{renormalized} \tilde{u}^{(\uparrow)}(p)=\lim_{\xi\to
\infty}Nu^{(\uparrow)}(p)=\left(
                                                 \begin{array}{c}
                                                   \cos{\frac{\theta}{2}} \\
                                                   0 \\
                                                   0 \\
                                                   \sin{\frac{\theta}{2}} \\
                                                 \end{array}
                                               \right), \;\;\;\;\;\;
                                               \tilde{u}^{(\downarrow)}(p)=\lim_{\xi\to
                                               \infty}Nu^{(\downarrow)}(p)=\left(
                                                     \begin{array}{c}
                                                       -\sin{\frac{\theta}{2}} \\
                                                       0 \\
                                                       0 \\
                                                       \cos{\frac{\theta}{2}} \\
                                                     \end{array}
                                                   \right).
\end{eqnarray}
Here, $N$ is the normalization factor and the upper tilde symbol is
used to indicate that these are normalized spinors in the $\xi\to
\infty$ limit. We see from Eq.(\ref{renormalized}) that the helicity
eigenstates which correspond to special spin orientation $\theta=0$,
converge to the chirality eigenstates in the $\xi\to \infty$ limit.
Indeed if we apply right and left chirality projection operators
$\hat R=\frac{1}{2}(1+\gamma_5)$ and $\hat
L=\frac{1}{2}(1-\gamma_5)$ to the helicity eigenstates we obtain
\begin{eqnarray}
\label{limit1}  \begin{array}{c}
                              \hat{R}\;\tilde u^{(+)}(p)=\tilde u^{(+)}(p) ,\;\;\;\;
                              \hat{R}\;\tilde u^{(-)}(p)=0\\
                              \hat{L}\;\tilde u^{(-)}(p)=\tilde u^{(-)}(p),\;\;\;\;
                              \hat{L}\;\tilde u^{(+)}(p)=0
                            \end{array}
\end{eqnarray}
Here we should remind that for special spin orientation $\theta=0$,
spin-up (spin-down) spinor corresponds to positive (negative)
helicity and instead of the notation "$\uparrow$" and "$\downarrow$"
we use "$+$" and "$-$". On the other hand this zero-mass behavior is
not valid for spinors with general spin orientation. If we apply the
chirality projection operators $\hat R$ and $\hat L$ to the spinors
(\ref{renormalized}) describing a general spin orientation we get
the following equalities which hold in the zero-mass limit:
\begin{eqnarray}
\label{limit2} \begin{array}{c} \hat R \tilde
u^{(\uparrow)}(p)=\cos{\left(\frac{\theta}{2}\right)}\;\tilde
u^{(+)}(p),\;\;\;\; \hat
R \tilde u^{(\downarrow)}(p)=-\sin{\left(\frac{\theta}{2}\right)}\;\tilde u^{(+)}(p)\\
\hat L\tilde
u^{(\uparrow)}(p)=\sin{\left(\frac{\theta}{2}\right)}\;\tilde
u^{(-)}(p),\;\;\;\; \hat L\tilde
u^{(\downarrow)}(p)=\cos{\left(\frac{\theta}{2}\right)}\;\tilde
u^{(-)}(p)
\end{array}
\end{eqnarray}
We see from Eq.(\ref{limit2}) that the spinors describing a general
spin orientation do not converge to one of the chirality eigenstates
left-handed or right-handed in the zero-mass limit. They are always
given by a mixed chirality. This fact can also be understood
intuitively from (\ref{movingspinors}) and (\ref{generalspinors1}).
We see from (\ref{movingspinors}) and (\ref{generalspinors1}) that
the operator $\exp{(-i\xi K_3)}$ tends to the projection onto the
first and last component of the spinor as the rapidity parameter
$\xi$ increases. In the limit $\xi\to \infty$ the second and third
components of $\exp{(-i\xi K_3)}u^{(\uparrow,\downarrow)}_{RF}$
disappear whereas the first and last component diverge. We can
regulate these divergent components through the normalization
procedure discussed in the previous page (see
Eq.(\ref{renormalized}) and the paragraph above). If the result of
the limit operator applied to a spinor of the form
(\ref{restspinors}) is to be an eigenvector of $\gamma_5$, then one
component of $\chi_\pm$ must be equal to 0. This is the case if and
only if $\theta=0$ or $\theta=\pi$.

The zero-mass behavior observed from (\ref{limit2}) is also evident
from Eqs.(\ref{spinup}) and (\ref{spindown}) where the spinors are
written explicitly as a mixed state composed of helicity
eigenstates. Since the helicity and chirality eigenstates coincide
in the zero-mass limit, this mixed helicity state represents a mixed
chirality state for $\xi\to \infty$. The amount of mixture is
determined by the angle $\theta$. For special orientation
$\theta=\pi/2$ (transverse polarization) we get the maximum mixing.
Here, we would like to draw the reader's attention to the fact that
{\it $\theta$ is not a dynamical variable}. It does not depend on
the rapidity parameter $\xi$. $\theta$ is the angle measured in the
frame in which the particle is at rest. Hence, one may call it
"proper angle" in analogy with the term proper time. Strictly
speaking the spinors given in (\ref{generalspinors1}) or
(\ref{generalspinors2}) describe a fermion which in its rest frame
has a spin orientation $\vec{n}=\sin\theta\; \hat x+ \cos\theta\;
\hat z$. Therefore the angle $\theta$ is not affected by
relativistic aberration.

It is well known that the Dirac equation decomposes into two Weyl
equations when mass is zero. Hence one may expect that the free
solutions of the Dirac equation converge to the solutions of one of
the Weyl equations in the $m\to0$ limit. Moreover the behavior of
the helicity eigenstates in the zero-mass limit is compatible with
this expectation. According to our opinion, the above arguments may
mislead some students or non-experts to think that the expectation
is valid in a general case. On the contrary as we have shown the
free spinor with general spin orientation cannot be described by one
of the Weyl equations in the zero-mass limit. We have deduced that
the spinor describing a general spin orientation is always given by
a mixed chirality state and thus both of the Weyl equations are
necessary simultaneously to describe the spinor with general spin
orientation even in the zero-mass limit.

At first glance, this odd zero-mass behavior of the general spinors
in (\ref{generalspinors1}) or (\ref{generalspinors2}) seems to
contradict the results of the little group analysis of Wigner
\cite{Wigner}. It is well known that massless particles are
described solely by pure helicity states and that the spin
orientations other than parallel or antiparallel to the direction of
momentum are not allowed for massless particles \cite{Wigner}.
Therefore it is very natural to expect that a spinor with a general
spin orientation gradually becomes more and more longitudinal and
its transverse component gradually vanishes as the mass goes to
zero. This behavior obviously contradicts to the one observed from
Eq.(\ref{limit2}). The crucial point which is generally skipped is
that the transition from general spin states to the helicity
eigenstates does not necessarily take place in a continuous manner.
If the mass is strictly zero then the particle moves at the speed of
light and consequently we cannot define its rest frame. Since it is
impossible to make a Lorentz transformation to a frame moving at the
speed of light, our calculations which leads to
Eqs.(\ref{generalspinors1}) and (\ref{generalspinors2}) are invalid
for strictly massless particles. On the other hand, if the fermion
has a nonzero mass we can make a Lorentz transformation to the rest
frame of the fermion. Hence, the general spinors given in
(\ref{generalspinors1}) or (\ref{generalspinors2}) describe the
fermions with mass greater than zero, i.e., $m>0$. Here we should
stress the fact that the general spinors describe massive fermions
no matter how small their masses are. In the point $m=0$ our
calculations and their results Eqs.(\ref{generalspinors1}) and
(\ref{generalspinors2}) become invalid. Since the limit $m\to0$ of a
free spinor $u^{(\uparrow,\downarrow)}(p)$ with general spin
orientation exists but not equal to one of the helicity eigenstates
$u^{(+,-)}(p)$, i.e., $\lim_{m\to0}u^{(\uparrow,\downarrow)}(p)\neq
u^{(+,-)}(p)$, we conclude that the free solutions of the Dirac
equation describing a general spin orientation have a discontinuity
at the point $m=0$. Therefore the zero-mass behavior observed from
Eq.(\ref{limit2}) does not contradict the results of \cite{Wigner}.
Strictly massless fermions are indeed described solely by helicity
eigenstates. However if the fermion has a non-zero mass then it is
allowed to have an arbitrary spin orientation which is different
from the momentum or opposite momentum direction. The transverse
polarization does not disappear in the $m\to0$ limit but it vanishes
instantly at the point $m=0$.

In closing, let us discuss some of the implications of this
zero-mass behavior of the Dirac spinors on particle physics.
Experimental results obtained in Super-Kamiokande and Sudbury
Neutrino Observatory verified the existence of neutrino oscillations
which shows that neutrinos have tiny but non-zero masses
\cite{Fukuda:1998mi,Ahmad:2001an}. Since the neutrino masses are
very tiny they are ultrarelativistic fermions at the energy scales
of current experiments. Therefore the results obtained in the
zero-mass limit can be applicable to neutrinos. In the Standard
Model of particle physics, neutrinos couple minimally to other
standard model particles only through the vertex
$\frac{1}{2}(1-\gamma_5)$  and hence the interaction project out the
"left" chiral component of the field and the "right" chiral
component decouples completely. Consequently, the cross section for
a neutrino with positive helicity is so tiny and can be neglected
for energies much greater than $m$. On the other hand as we have
deduced the spinor with general spin orientation is always given by
a mixed chirality state determined by the non-dynamical variable
$\theta$. Hence, the weak interaction does not annihilate the
fermion field describing a general spin orientation in the $m\to0$
limit. Eventually, although the cross section for a neutrino with
positive helicity goes to zero as $m\to0$, the cross section remains
finite for a neutrino with general spin orientation
\cite{Sahin:2015ofl}. Specifically if the neutrino is transversely
polarized ($\theta=\pi/2$) then the polarized cross section is
almost half of the cross section for negative helicity
neutrino.\footnote{Here we should note that the neutrinos interact
in flavor eigenstates which are given by a superposition of the mass
eigenstates. Hence, in the reality the calculation of the cross
section is more complicated than we have discussed in this paper.}
The above results imply that the production or the absorption
probability of the neutrinos with arbitrary spin orientation through
Standard Model reactions cannot be neglected in general.

\end{document}